\documentstyle[aps,preprint,tighten]{revtex}

\begin{document}
\draft


\title{$f_K/f_\pi$ ratio from QCD sum rules}
\author{
  Seungho Choe\thanks{E-mail: schoe@physics.adelaide.edu.au}
\footnote{Present Address : Special Research Centre for the
  Subatomic Structure of Matter, University of Adelaide,
  Adelaide, SA 5005, Australia}
 and
Su Houng Lee\thanks{E-mail: suhoung@phya.yonsei.ac.kr}}
\address{Department of Physics\\
Yonsei University, Seoul 120--749, Korea}
\maketitle
\begin{abstract}

Using the correlation function of the 
axial vector mesons,
we present a QCD sum rule calculation for the 
 decay constants $f_\pi$ and $f_K$. 
Our calculations are only weakly dependent on the 
SU(3) breaking-parameter for the QCD vacuum  and give the 
ratio $f_K/f_\pi$ = 1.11 $\pm$ 0.02.

\end{abstract}
\vspace{1cm}

\section{Introduction}

Understanding the explicit SU(3) symmetry breaking effect in physical 
quantities, such as mass slitting, coupling constants, and decay constants, 
has been a subject of research in models of QCD for many years.   
Among those models, the QCD sum rule\cite{SVZ} provides 
a semi-direct calculation of 
QCD in that it relates, via the Borel transformed dispersion relation, 
the physical quantities to perturbative QCD supplemented by
the non-perturbative nature of the QCD vacuum summarized systematically in 
non-vanishing condensates.    Therefore, the main SU(3)-breaking effects are  
included systematically in perturbative quark-mass corrections
(i.e., $m_u = m_d \neq m_s$)
and in the different quark condensates ($\langle\bar{u}u\rangle =
\langle\bar{d}d\rangle \neq \langle\bar{s}s\rangle$).
Using these prescriptions, the mass splittings within meson and baryon
multiplets were calculated and it was found that the best fit was 
obtained with $m_s \sim 150 $ MeV and $\gamma= \langle \bar{s} s \rangle
 / \langle \bar{u} u \rangle -1 \sim - 0.2$, where we assumed
 $ m_u =m_d \sim 0$ and $\langle \bar{u} 
 u \rangle=\langle \bar{d} d \rangle$\cite{RRY}.  

Despite all this success, it is not always possible to calculate all 
the physical quantities in the QCD sum rules, especially those related to 
Goldstone bosons.  There are two reasons for this.  First, the QCD sum rule is
based on the operator product expansion (OPE) 
for which convergence is guaranteed 
only for large space-like 
momenta.  This means, it is rather difficult to obtain information about
the quantities carried by light Goldstone bosons 
for which information at rather
small momentum transfer is needed.  
Second, for pseudo-scalar pseudo-scalar correlation functions, direct
instantons can contribute, which  spoils the convergence of the OPE.  
Therefore, it is not possible to calculate, for example, 
the pseudo-scalar meson masses in the QCD sum rules.   
Nonetheless, using suitable methods, it was found that  
low energy theorems, such as the Gell-Mann, Oakes, Renner relation\cite{GOR}
and the Goldberger-Treiman relation\cite{GT}, could be
derived or seen to hold within QCD sum rule approaches (see Appendices).
Recently, it was shown further that even the chiral log behavior
could be put in consistently by suitably modifying the continuum
part\cite{LCCG95}.   
These successes imply that by appropriately choosing the correlation
function and improving the continuum part, we can estimate the effects of
explicit chiral symmetry breaking, even for quantities related to the 
Goldstone bosons\footnote{As an example, in Ref. \cite{CCL96},
$g_{K N \Lambda}$ and $g_{K N \Sigma}$ are calculated, and
they are compared to $g_{\pi NN}$.}.

In this work, we proceed along these line by presenting 
a QCD sum-rule calculation for the 
 decay constants $f_\pi$ and $f_K$ and their ratio 
by using the correlation function of the axial vector currents for which 
no contamination from direct instantons is expected
\cite{Shuryak}.    
Our calculation for the ratio gives $f_K/f_\pi$ = 1.11 $\pm$ 0.02.
There have been many different models
\cite{Ch,Do,La,Go,Ha,Kr,God,Ko,Sa,Mc,Mu,Le,Ho,De,Ham,Ab,Bu,Kl,Hat,Al,Re}
to get $f_\pi$, $f_K$ and $f_K/f_\pi$. Compared to those 
calculations, our value for the ratio lies at the lower end.  
However,
the present calculations are only weakly dependent on the 
SU(3)-breaking parameter for the QCD vacuum and give values
close to those of a recent
lattice calculation.

In Sec. II, we present mass formula for the two axial vector mesons
($a_1$ (non-strange axial vector meson) and $K_a$ (strange axial vector
meson)) by using the QCD sum rules.
In Sec. III, we obtain the decay constants $f_\pi$ and $f_K$ and the couplings
$4\pi/g_{a_1}^2$ and $4\pi/g_{K_a}^2$ from those sum rules which contain
 two SU(3) symmetry-breaking parameters, $m_s$ and $\gamma$.
We summarize our results in Sec. IV.

\section{Sum Rules for $a_1$ and $K_a$}{\label{a1k1}}

Consider the T--product of the axial vector currents \cite{SVZ,RYR,Gov}:
\begin{eqnarray}
\Pi_{\mu\nu} (q^2) &=& i \int d^4x e^{iqx}\langle
      T ( J_\mu^A(x) J_\nu^A(0) )\rangle ,
\end{eqnarray}
 where $J_\mu^A(x) = \bar{q}(x) \gamma_\mu \gamma_5 q(x)$ and the $q's$ 
 are either  $u$ or $d$ quarks only.
 Then, this current couples to the $a_1$ (1$^{++}$)
 and $\pi$ (0$^{-+}$) mesons, and $\Pi_{\mu\nu} (q^2)$  above can be
 decomposed as follows:
\begin{eqnarray}
\Pi_{\mu\nu} (q^2) = ({q_\mu q_\nu \over q^2} - g_{\mu\nu}) \Pi_A (q^2)
			 + {q_\mu q_\nu \over q^2} \Pi_P (q^2) ,
\end{eqnarray}
where the imaginary parts of  $\Pi_A (q^2)$ and $\Pi_P (q^2)$ receive 
contributions from $1^{++}$ and $0^{-+}$ states, respectively.
One can extend this argument to  currents involving the s-quark.
For example, we can take the current
$J_\mu^A(x) = \bar{u}(x) \gamma_\mu \gamma_5 d(x)$ for the $a_1$ meson
 and $\bar{u}(x) \gamma_\mu \gamma_5 s(x)$ for the $K_a$ meson, where
 $K_a$ is the chiral partner of $a_1$.
 In the SU(3) symmetric limit the $^3P_1$ and the $^1P_1$ states do not mix,
 just like the $a_1$ and $b_1$ mesons. However, for the s-quark
 mass greater than the u and the d-quark masses,  
the $^3P_1$ and the $^1P_1$ states mix
 to give the physical $K_1$ ($K_1$(1270) and $K_1$(1400))
 states\cite{Ca,To,Ma,Gov,Cap,Li,Su,Bl,Bi}.

 $\Pi_{\mu\nu} (q^2)$ can also be written as 
\begin{eqnarray}
\Pi_{\mu\nu} = -\Pi_1 (q^2)~g_{\mu\nu} + \Pi_2 (q^2)~q_\mu q_\nu .
\end{eqnarray}
We get $f_\pi$ and $f_K$ from  $\Pi_2 (q^2)$.
On the OPE side, after the Borel transformation, we obtain the following
for the $K_a$ meson to the leading order in $\alpha_s$:
\begin{eqnarray}
\frac{1}{\pi} \int e^{-s/M^2} Im \Pi_2(s) ds  = \frac{1}{4\pi^2} M^2
    [ 1 + \frac{\alpha_s}{\pi} - \frac{3}{M^2}(m_u^2 + m_s^2)
     + \frac{\pi^2}{3M^4} \langle\frac{\alpha_s}{\pi} G^2\rangle
\nonumber \\
     + \frac{4\pi^2}{M^4} (m_u \langle\bar{u}u\rangle
     + m_s \langle\bar{s}s\rangle)
     + \frac{64 \pi^3 \alpha_s}{81 M^6}
     (\langle\bar{q}q\rangle^2 + \langle\bar{s}s\rangle^2
     + 9\langle\bar{q}q\rangle\langle\bar{s}s\rangle)] ,
\label{ope2}
\end{eqnarray}
where we have  
assumed $\langle\bar{u}u\rangle$ =
$\langle\bar{d}d\rangle$ $\equiv$ $\langle\bar{q}q\rangle$.
In the following we neglect the terms proportional to $m_u$
(and $m_d$ for the case of $a_1$) because they give
corrections of less than 0.2\% to the OPE side at the relevant Borel region.
 For the four quark-condensates,
we assume the vacuum saturation hypothesis, i.e.,
\begin{eqnarray}
\langle \bar{q} \Gamma_i \frac{\lambda^a}{2} q \bar{q} \Gamma_i
\frac{\lambda^a}{2} q \rangle = - \frac{1}{N^2} Tr (\Gamma_i \Gamma_i)
Tr (\frac{\lambda^a}{2} \frac{\lambda^a}{2}) \langle\bar{q}q\rangle^2
\end{eqnarray}
with N=12 being a normalization factor.
For example,
\begin{eqnarray}
\langle \bar{q} \gamma_\mu \lambda^a q \bar{q} \gamma_\mu \lambda^a q \rangle
&=& - \frac{16}{9} \langle\bar{q}q\rangle^2 ,
 \nonumber \\
\langle \bar{q} \gamma_5 \lambda^a q \bar{q} \gamma_5 \lambda^a q \rangle
&=& - \frac{4}{9} \langle\bar{q}q\rangle^2 .
\end{eqnarray}
This approximation has been criticized by some people \cite{GBP}.
However, it turns out to be adequate in the case of
the $a_1$ sum rules \cite{Gov}. 
This is also true for the case of the $K_a$ sum rules
because $K_a$ has the same structure.
For the $a_1$, we let $m_s$=0 and replace
$\langle\bar{s}s\rangle$ with $\langle\bar{q}q\rangle$ above.
On the phenomenological side, the spectral density $Im \Pi_{2}$
can be expressed as follows
(a pole contribution plus a continuum contribution):
\begin{eqnarray}
\frac{1}{\pi} Im \Pi_2 &=& f_{\pi,K}^2 \delta(s-m_{\pi,K}^2)
      + \frac{m_{a_1,K_a}^2}{g_{a_1,K_a}^2} \delta(s-m_{a_1,K_a}^2)
\nonumber\\*
      &+& \frac{1}{4\pi^2} (1 + \frac{\alpha_s}{\pi}) \theta(s-s_0) ,
\label{phe2}
\end{eqnarray}
where the constants $f_\pi$, $f_K$, $g_{a_1}$, and $g_{K_a}$ are
defined in the usual way:
\begin{eqnarray}
\langle0| \bar{u}\gamma_\mu\gamma_5 d|\pi\rangle = i f_\pi p_\mu,
~~\langle0| \bar{u}\gamma_\mu\gamma_5 d|a_1\rangle =
                   \frac{m_{a_1}^2}{g_{a_1}} \epsilon_\mu ,
\end{eqnarray}
and
\begin{eqnarray}
\langle0| \bar{u}\gamma_\mu\gamma_5 s|K\rangle = i f_K p_\mu,
~~\langle0| \bar{u}\gamma_\mu\gamma_5 s|K_a\rangle =
		   \frac{m_{K_a}^2}{g_{K_a}} \epsilon_\mu .
\end{eqnarray}
$m_{a_1}$ and $m_{K_a}$ represent the masses of the $a_1$ and the $K_a$
mesons, respectively, and $s_0$ is the continuum threshold.

After inserting Eq. (\ref{phe2}) into
Eq. (\ref{ope2}), we can get the
following expression: 
\begin{eqnarray}
f_{\pi,K}^2 e^{-\frac{m_{\pi,K}^2}{M^2}}
&+& {m_{a_1,K_a}^2 \over g_{a_1,K_a}^2} e^{-\frac{m_{a_1,K_a}^2}{M^2}}
\nonumber\\*
&=& {1 \over 4\pi^2} M^2 [(1+\frac{\alpha_s}{\pi})(1-e^{\frac{s_0}{M^2}})
+\frac{A}{M^2}
    +\frac{B}{M^4}+\frac{C}{M^6}] ,
\label{form2}
\end{eqnarray}
where for $K_a$,
\begin{eqnarray}
   A &=& - 3 m_s^2 , \nonumber\\
   B &=& \frac{\pi^2}{3} \langle\frac{\alpha_s}{\pi} G^2\rangle
   + 4\pi^2 m_s \langle\bar{s}s\rangle ,\nonumber\\
   C &=& \frac{64}{81} \pi^3 \alpha_s (\langle\bar{q}q\rangle^2
 + \langle\bar{s}s\rangle^2 + 9 \langle\bar{q}q\rangle\langle\bar{s}s\rangle) .
\label{const2}
\end{eqnarray}
 For the $a_1$ meson, we put $m_s$=0 and replace $\langle\bar{s}s\rangle$
by $\langle\bar{q}q\rangle$
in Eq. (\ref{const2}).

\section{$f_\pi$, $f_K$, and $f_K/f_\pi$}{\label{fpifk}}

In our formula (Eq. (\ref{form2})), there are several
sources of uncertainties in the OPE: these  are the magnitude of the s-quark 
condensate, contributions of higher-dimensional
operators, the effect of the running coupling constants $\alpha_s$ (M).
We examine $f_\pi$, $f_K$ and the 
ratio $f_K/f_\pi$ for four cases and find their changes case by case.
These four cases are used to estimate the uncertainties on the OPE side.

\subsection{Case I : $\langle\bar{s}s\rangle$=0.8$\langle\bar{q}q\rangle$}

 From the expression in
Eq. (\ref{form2}), we can determine
$f_\pi$ (or $f_K$), $s_0$, and the coupling
$4\pi/g_{a_1}^2$ (or $4\pi/g_{K_a}^2$)
by using the experimentally known $a_1$ (or $K_a$) mass.
We use $m_{a_1}$ = 1.230 $\pm$ 0.040 GeV \cite{PRD}
and $m_{K_a} \sim$ 1.340 GeV \cite{Gov,Su,Bl,Bi}.

We have to determine $f_{\pi,K}$, $g_{a_1,K_a}^2$, $s_0$, and
the Borel interval.
To do this, we use a best-fit method. The equation has the following form:
\begin{eqnarray}
C_1 ~g_1 (M^2) + C_2 ~g_2 (M^2) = g_3 (M^2) ,
\label{mini1}
\end{eqnarray}
where $C_1$ = $f_{\pi,K}^2$ and $C_2$ = $1/g_{a_1,K_1}^2$.
We want to determine $C_1$ and $C_2$ by 
 minimizing  $(C_1 ~g_1 + C_2 ~g_2 - g_3)^2$
with a fixed $s_0$ and an appropriate Borel interval:
\begin{eqnarray}
\int_{M_i^2}^{M_f^2} (C_1 ~g_1 + C_2 ~g_2 - g_3)^2 ~dM^2 = {\rm minimum} .
\end{eqnarray}
The Borel interval $M^2$ is restricted by the following conditions:
OPE convergence and pole dominance.
The lower limit of $M^2$, $M_i^2$ is determined as the value at which the
contribution of the power corrections on the OPE side is less than 30\%.
The upper limit $M_f^2$ is
determined as the value where the continuum contribution
on the right-hand sides of (Eq. (\ref{form2})) is less than 50$\%$ of the  
total.
 After determining 
$f_{\pi,K}$, $g_{a_1,K_a}^2$, we repeat the procedure with a different 
threshold $s_0$ until the variation 
 $h(M^2) \equiv ~C_1 ~g_1 (M^2) + C_2 ~g_2 (M^2) - g_3 (M^2)$ is minimized.   
When the variation is the least, we take those values of
$f_{\pi,K}$, $g_{a_1,K_a}^2$, $s_0$ and the Borel interval as our results.

In Tables \ref{table_d1} and \ref{table_d2} we summarize our results.
We choose two values for the quark condensate,
$\langle\bar{q}q\rangle$ = --(0.230 GeV)$^3$ and
$\langle\bar{q}q\rangle$ = --(0.250 GeV)$^3$.
Table \ref{table_d1} is the result for the case of
$\langle\bar{q}q\rangle$= --(0.230 GeV)$^3$,
and Table \ref{table_d2} is that for the case of
$\langle\bar{q}q\rangle$= --(0.250 GeV)$^3$.
 Throughout this paper,
we take $\langle\bar{s}s\rangle$ =
0.8 $\langle\bar{q}q\rangle$, 
$\langle\frac{\alpha_s}{\pi} G^2\rangle$ = 0.015 GeV$^4$, and
$\alpha_s$ = 0.5.
If we take
$\langle\bar{s}s\rangle$ = 0.6 $\langle\bar{u}u\rangle$, we obtain
$f_K$ = 0.146 GeV and $\frac{4\pi}{g_{K_a}^2}$ = 0.411
for $\langle\bar{q}q\rangle$= --(0.230 GeV)$^3$.
For the case of
$\langle\bar{s}s\rangle$ = 1.0 $\langle\bar{u}u\rangle$,
$f_K$ = 0.142 GeV and $\frac{4\pi}{g_{K_a}^2}$ = 0.429.
Overall,  the value of $f_K$  is not sensitive (within  a few percent)  to 
any variations in the $u$ and the $d$ or the $s$-quark condensate within its 
expected range.  

\subsection{Case II : $\langle\bar{s}s\rangle$=
0.8$\langle\bar{q}q\rangle$ and Including Mixed Condensate}

Here, we include the contributions from the dimension-6 mixed condensates,  
$m_s \langle g_s \bar{s} ~\sigma \cdot G ~s\rangle$ for $K_a$,
where $\sigma \cdot G \equiv \sigma_{\mu\nu} \cdot G^{\mu\nu}$
and $\sigma_{\mu\nu} = \frac{i}{2} [ \gamma_\mu , \gamma_\nu]$, and 
 examine the changes.
If the mixed condensate is included, the constant C in
Eq. (\ref{const2}) is changed as follows: 
C$^\prime$ = C - $\frac{8\pi^2}{15} m_s \langle\bar{s}s\rangle$,
where we let $\langle g_s \bar{s} ~\sigma \cdot G ~s\rangle$
$\equiv$ 2 $m_0^2 \langle\bar{s}s\rangle$ =
0.8$\langle\bar{s}s\rangle$\cite{Ioffe83}.
The value $m_0^2$ is quoted
from the standard QCD sum-rule estimation, $\sim$ 0.4 GeV$^2$ \cite{Be,Ovc}.
The results are in Tables \ref{table_d3} and \ref{table_d4}.
Of course, there are no changes in $f_\pi$ and $\frac{4\pi}{g_{a_1}^2}$,
but
there are small changes in $f_K$ and the coupling
$\frac{4\pi}{g_{K_a}^2}$.
A somewhat larger value of $m_0^2$ has been obtained from different approaches
\cite{KS,Shu,BR,MR}.
However even with that value, the change in $f_K$ is very small.
If we take the value suggested in Ref. \cite{MR}, $m_0^2 \sim$ 1.2 GeV$^2$,
we get $f_K$ = 0.149 GeV and $\frac{4\pi}{g_{K_a}^2}$ = 0.376 for
$\langle\bar{q}q\rangle$= --(0.230 GeV)$^3$.
For $\langle\bar{q}q\rangle$= --(0.250 GeV)$^3$, we have
$f_K$ = 0.150 GeV and $\frac{4\pi}{g_{K_a}^2}$ = 0.365.

\subsection{Case III : $\langle\bar{s}s\rangle$=
0.8$\langle\bar{q}q\rangle$ with Variation of $\alpha_s$(M)}

We now take into account the running coupling constant $\alpha_s$(M).
We use the forms used in Ref. \cite{Gov} :
\begin{eqnarray}
\alpha_s (M) = \frac{4\pi}{9} \frac{1}{ln(M^2/\Lambda^2)}
\end{eqnarray}
with $\Lambda$ $\simeq$ 150 MeV and
\begin{eqnarray}
\langle\bar{q}q (M) \rangle =
\left( \frac{\alpha_s (\mu)}{\alpha_s (M)} \right)^{4/9}
                      ~\langle\bar{q}q (\mu) \rangle ,
\end{eqnarray}
where $\mu$ is a normalization scale.
For the four quark-condensates,
\begin{eqnarray}
\alpha_s(M) ~\langle\bar{q}q(M)\rangle ^2
= \alpha_s(M) ~\left( \frac{\alpha_s (\mu)}{\alpha_s (M)} \right)^{8/9}
            ~\langle\bar{q}q (\mu) \rangle ^2
\simeq \alpha_s (\mu) ~\langle\bar{q}q (\mu) \rangle^2 ,
\end{eqnarray}
where we have assumed the vacuum saturation hypothesis, as before, and
$\alpha_s(\mu)$ = 0.5.
Using this, we obtain new results from Eq. (\ref{form2}), which are given in 
Tables \ref{table_d5} and \ref{table_d6}.
As one can see, the decay constants $f_\pi$ and $f_K$ and the 
two couplings are
smaller than before. However, the change is only within a few percent.
Our results are not sensitive to the
choice of $\alpha_s (\mu)$. If we take another
$\alpha_s (\mu)$, i.e.,
0.6 as usual, the change is less than 1\% for the decay constants and 3\% for
the couplings.

\subsection{Case IV : $\langle\bar{s}s\rangle$=
0.8$\langle\bar{q}q\rangle$ and Including Mixed Condensate with
Variation of $\alpha_s$(M)}

In this case, we include the mixed condensate considered in Case III,
and use the running
coupling constant $\alpha_s$(M).
The anomalous dimension of the quark-gluon mixed operator
$g_s \bar{q} \sigma \cdot G q $ is small and can be neglected \cite{Shi}.
The new results are in Table \ref{table_d7} and \ref{table_d8}.
These are our final results.
Comparing our values to the experimental values ($f_\pi$ $\sim$ 0.131 GeV, and
$f_K$ $\sim$ 0.160 GeV) \cite{PRD},
$f_\pi$ is very similar, while $f_K$ is smaller than  the experimental
value.
With $\langle\bar{q}q\rangle$ = --(0.230 GeV)$^3$, we get
$f_\pi$ = 0.130 $\pm$ 0.002 GeV and $f_K$ = 0.144 GeV.
The couplings are $4 \pi/g_{a_1}^2$ = 0.42 $\pm$ 0.02
and $4 \pi/g_{K_a}^2$ = 0.38.
With $\langle\bar{q}q\rangle$ = --(0.250 GeV)$^3$,
$f_\pi$ = 0.132 $\pm$ 0.003 GeV and $f_K$ = 0.144 GeV.
Their couplings
$4 \pi/g_{a_1}^2$ and $4 \pi/g_{K_a}^2$
are 0.40 $\pm$ 0.02 and 0.38, respectively.
The error bars for $f_\pi$ and 4$\pi/g_{a_1}^2$ come from the uncertainty
in the  $a_1$ mass, $m_{a_1} = 1.230 \pm 0.040$ GeV\cite{PRD}.

\section{Discussion}{\label{discuss}}

In Tables \ref{table_d9} and \ref{table_d10} we summarize the ratio
$f_K/f_\pi$ for the four cases considered.  
One can see that there is not much of a difference in the ratio between
those cases.
This indicates that the uncertainties coming from the OPE side
in our formula are very small.
For the sake of reference, we compare our ratio with those of other
models in Table \ref{table_d11}.
One can see that our result 1.11 $\pm$ 0.02
for $\langle\bar{q}q\rangle$ = --(0.230 GeV)$^3$
is very similar to
that from lattice gauge theory, but smaller than the
experimental value, 1.22 $\pm$ 0.02. However, the error is within 10$\%$.
It should be noted that some models in the Table
predict a ratio which is similar to the experimental
value, but do not give the  correct $f_\pi$ and $f_K$.

We also check that a different $K_a$ mass doesn't change our result very much.
For example, if we take the mass as $K_a$ = 1.270 GeV (the same value as
that of $K_1$ (1270)), $f_K$ = 0.140 GeV.
If we take the mass as $K_a$ = 1.400 GeV (that of $K_1$ (1400)),
$f_K$ = 0.148 GeV.
Among the parameters, the ratio is most sensitive to $m_s$.
In the case with $\langle\bar{q}q\rangle$ = --(0.230 GeV)$^3$
, if we let $m_s$ = 0.125 GeV, we get
$f_K$=0.148 GeV and $f_K/f_\pi$ = 1.14$\pm$0.02. In the case of
$m_s$=0.175 GeV, $f_K$=0.140 and $f_K/f_\pi$= 1.08$\pm$0.02. 
Thus, a better determination of $m_s$ will pin down the value of $f_K$ 
with better accuracy.  
It seems contradictory at first that the result depends more sensitively on  
 $m_s$ than on the mass of $K_a$.   However, it should be noted that the 
mass of $K_a$ comes dominantly from chiral symmetry breaking, and the 
explicit SU(3) symmetry-breaking effect is rather small.  This is evident 
from comparing its mass to that of the $a_1(1230)$ meson.
On the other hand, in order to extract decay constants from experiments,
we have to determine some elements of
the CKM (Cabbibo-Kobayashi-Maskawa) matrix,
such as $|V_{ud}|$ and $|V_{us}|$ for $f_\pi$ and $f_K$, respectively.
These elements are also closely related to the current quark masses, 
 $m_u, m_d$, and $m_s$.

\acknowledgements

The authors thank H. Weigel for useful comments on the results of 
the NJL model
calculation.
The work of SHL and SC were supported in part by the Basic Science
Research
Institute Program, Ministry of Education of Korea, no.
BSRI-97-2425 and by the Korea Science and Engineering Foundation
 through grant no. 971-0204-017-2 and 
through the Center for Theoretical Physics at Seoul National University.
One of the authors (SC) wishes to acknowledge the financial support of 
the Korea Research Foundation (KRF) made in the program year 1997.
The work of SC was supported in part by
the Special Research Centre for the Subatomic Structure of Matter at
the University of Adelaide where it is completed.

\appendix 
\section{}
Here, we derive the  Goldberger-Trieman relation within the QCD sum rule
approach.
Consider the following nucleon correlation function: 
\begin{eqnarray}
\Pi(q;\pi(p))= i \int d^4x e^{iqx} \langle 0| T[\eta(x) \bar{\eta}(0)]
 |\pi(p) \rangle ,
\end{eqnarray}
where $\pi(p)$ is a pion with momentum $p$, and $\eta$ is a nucleon
interpolating field without any derivative.  
The nucleon interpolating field transforms as follows under the 
 SU(2) axial rotation:
\begin{eqnarray}
\label{ap2}
[ Q_5^a,\eta]=-\gamma_5 \frac{\tau^a}{2} \eta,
\end{eqnarray}
where $Q_5^a$ is the axial charge.  In  the soft-pion limit,
the OPE side is the commutator with the axial charge and, by using the 
relation in Eq.(\ref{ap2}), can be shown to be 
\begin{eqnarray}
\label{apope1}
\Pi^{{\rm OPE}}(q;\pi)=\frac{2}{f_\pi} \Pi^{{\rm OPE}}_1(q^2) \gamma_5,
\end{eqnarray}
where $\Pi^{{\rm OPE}}_1(q^2)$ is the OPE side of the following part of the
nucleon-correlation function in vacuum:
\begin{eqnarray}
\Pi(q;0) \equiv \Pi_1(q^2) +\Pi_q(q^2) \gamma^\mu q_\mu .
\end{eqnarray}

As for the phenomenological side, we assume an interaction lagrangian of the 
pion  and the nucleon, ${\cal L}_I=g_{\pi N} \bar{N} \gamma_5 \pi N$. 
Then, we have,
\begin{eqnarray}
\label{apphen1}
\Pi^{{\rm Phen}}(q;\pi)=\frac{\lambda_N^2 g_{\pi N} \gamma_5}{q^2-m_N^2}
+\frac{2}{f_\pi}\Pi^{cont}_1(q^2) \gamma_5.
\end{eqnarray}
We also have the following vacuum sum rule:
\begin{eqnarray}
\Pi_1^{{\rm OPE}}(q^2)=\frac{\lambda_N^2 m_N}{q^2-m_N^2}+
\Pi_1^{cont}(q^2).
\end{eqnarray}
Using this and
comparing Eqs. (\ref{apope1}) and (\ref{apphen1}), we have 
the Goldberger-Trieman relation
\begin{eqnarray}
g_{\pi N}=\frac{2m_N}{f_\pi} .
\end{eqnarray}
%
\section{}
Here, we show that the pion hadron T matrix $T_{\pi-H} \rightarrow 0$ (Adler 
Zero) in the chiral limit and the pion momentum $p \rightarrow 0$. 
Consider the correlation function 
\begin{eqnarray}
V(q;2\pi(p))= i \int d^4x e^{iqx} \langle \pi(p)| T[H(x) \bar{H}(0)]
 |\pi(p) \rangle .
\end{eqnarray}
In the  soft-pion limit, using the commutation relation with the axial
current twice, 
the OPE side  can be shown to be 
\begin{eqnarray}
\label{apope2}
V^{{\rm OPE}}(q;2\pi)=-\frac{c_1}{f_\pi^2} V^{{\rm OPE}}_1(q,0) 
+ \frac{c_1}{f_\pi^2} A^{{\rm OPE}}_1(q,0),
\end{eqnarray}
where $A^{{\rm OPE}}_1(q,0)$ is the correlation function between the
axial partner of the hadronic current $H$.

As for the phenomenological side, there will be a double pole, whose 
residue is the T matrix:
\begin{eqnarray}
\label{apphen2}
V^{{\rm Phen}}(q;2\pi)=
\frac{T}{(q^2-m_H^2)^2}+V^{pole}(q;2\pi)+V^{cont}(q;2\pi)
+\frac{c_1}{f\pi^2}[A^{pole}(q;0)+A^{cont}(q;0)] .
\end{eqnarray}
We note
\begin{eqnarray}
V^{pole}(q;2\pi) \stackrel{q\rightarrow0}{\longrightarrow}
-\frac{c_1}{f_\pi^2} V^{pole}(q;0), \nonumber \\
V^{cont}(q;2\pi) \stackrel{q\rightarrow0}{\longrightarrow}
-\frac{c_1}{f_\pi^2} V^{cont}(q;0).
\end{eqnarray}
Using the sum rules
\begin{eqnarray}
V(q;0)^{{\rm OPE}}&=&V(q;0)^{pole}+ V(q;0)^{cont.}, 
\nonumber \\ 
A(q;0)^{{\rm OPE}}&=&A(q;0)^{pole}+ A(q;0)^{cont.}
\end{eqnarray}
and comparing Eqs. (\ref{apope2}) and (\ref{apphen2}), we have
\begin{eqnarray}
T=0
\end{eqnarray}
in the chiral limit with zero incoming four momentum.


\newpage
\begin{table}
\caption{Case I : $\langle\bar{q}q\rangle$~=
~--(0.230 GeV)$^3$, $\langle\bar{s}s\rangle$ = 0.8 $\langle\bar{q}q\rangle$.}
\label{table_d1}
\begin{center}
\begin{tabular}{ c c c c c c}
axial meson & mass (GeV) & $s_0$ (GeV$^2$) & $M^2$ (GeV$^2$) &
 $f_\pi$, $f_K$ (GeV) &  ${4\pi \over g_{a_1, K_a}^2}$ \\
\hline
$a_1$   &  1.190 &  2.15  & 1.35--3.10 & 0.130 & 0.410 \\
        &  1.230 &  2.35  & 1.35--3.40 & 0.132 & 0.427 \\
        &  1.270 &  2.55  & 1.35--3.65 & 0.134 & 0.441 \\
\hline
$K_a$   &  1.340 &  2.85  & 0.85--4.00 & 0.143 & 0.425  \\
\end{tabular}
\end{center}
\end{table}

\begin{table}
\caption{Case I : $\langle\bar{q}q\rangle$~=
~--(0.250 GeV)$^3$, $\langle\bar{s}s\rangle$ = 0.8 $\langle\bar{q}q\rangle$.}
\label{table_d2}
\begin{center}
\begin{tabular}{ c c c c c c}
axial meson & mass (GeV) & $s_0$ (GeV$^2$) & $M^2$ (GeV$^2$) &
 $f_\pi$, $f_K$ (GeV) &  ${4\pi \over g_{a_1, K_a}^2}$ \\
\hline
$a_1$   &  1.190 &  2.10  & 2.25--3.05 & 0.132 & 0.392 \\
        &  1.230 &  2.25  & 2.25--3.25 & 0.135 & 0.397 \\
        &  1.270 &  2.45  & 2.25--3.55 & 0.137 & 0.413 \\
\hline
$K_a$   &  1.340 &  2.80  & 1.00--3.90 & 0.143 & 0.417  \\
\end{tabular}
\end{center}
\end{table}

\newpage
\begin{table}
\caption{Case II : $\langle\bar{q}q\rangle$~=~--(0.230 GeV)$^3$,
$\langle\bar{s}s\rangle$ = 0.8 $\langle\bar{q}q\rangle$,
and a including mixed condensate.}
\label{table_d3}
\begin{center}
\begin{tabular}{ c c c c c c}
axial meson & mass (GeV)& $s_0$ (GeV$^2$) & $M^2$ (GeV$^2$) &
 $f_\pi$, $f_K$ (GeV) &  ${4\pi \over g_{a_1, K_a}^2}$ \\
\hline
$a_1$   &  1.190 &  2.15  & 1.35--3.10 & 0.130 & 0.410 \\
        &  1.230 &  2.35  & 1.35--3.40 & 0.132 & 0.427 \\
        &  1.270 &  2.55  & 1.35--3.65 & 0.134 & 0.441 \\
\hline
$K_a$   &  1.340 &  2.75  & 1.05--3.85 & 0.146 & 0.402  \\
\end{tabular}
\end{center}
\end{table}

\begin{table}
\caption{Case II : $\langle\bar{q}q\rangle$~=~--(0.250 GeV)$^3$,
$\langle\bar{s}s\rangle$ = 0.8 $\langle\bar{q}q\rangle$,
and including a mixed condensate.}
\label{table_d4}
\begin{center}
\begin{tabular}{ c c c c c c}
axial meson & mass (GeV)& $s_0$ (GeV$^2$) & $M^2$ (GeV$^2$) &
 $f_\pi$, $f_K$ (GeV) &  ${4\pi \over g_{a_1, K_a}^2}$ \\
\hline
$a_1$   &  1.190 &  2.10  & 2.25--3.05 & 0.132 & 0.392 \\
        &  1.230 &  2.30  & 2.25--3.30 & 0.134 & 0.411 \\
        &  1.270 &  2.45  & 2.25--3.55 & 0.137 & 0.413 \\
\hline
$K_a$   &  1.340 &  2.70  & 1.20--3.80 & 0.146 & 0.392  \\
\end{tabular}
\end{center}
\end{table}

\newpage
\begin{table}
\caption{Case III : $\langle\bar{q}q\rangle$~=~--(0.230 GeV)$^3$ and 
$\langle\bar{s}s\rangle$ = 0.8 $\langle\bar{q}q\rangle$
with variation of $\alpha_s$(M).}
\label{table_d5}
\begin{center}
\begin{tabular}{ c c c c c c}
axial meson & mass (GeV)& $s_0$ (GeV$^2$) & $M^2$ (GeV$^2$) &
 $f_\pi$, $f_K$ (GeV) &  ${4\pi \over g_{a_1, K_a}^2}$ \\
\hline
$a_1$   &  1.190 &  2.30  & 1.35--3.30 & 0.128 & 0.406 \\
        &  1.230 &  2.50  & 1.35--3.60 & 0.130 & 0.420 \\
        &  1.270 &  2.70  & 1.35--3.90 & 0.132 & 0.431 \\
\hline
$K_a$   &  1.340 &  2.95  & 0.85--4.15 & 0.142 & 0.403  \\
\end{tabular}
\end{center}
\end{table}

\begin{table}
\caption{Case III : $\langle\bar{q}q\rangle$~=~--(0.250 GeV)$^3$ and
$\langle\bar{s}s\rangle$ = 0.8 $\langle\bar{q}q\rangle$
with variation of $\alpha_s$(M).}
\label{table_d6}
\begin{center}
\begin{tabular}{ c c c c c c}
axial meson & mass (GeV)& $s_0$ (GeV$^2$) & $M^2$ (GeV$^2$) &
 $f_\pi$, $f_K$ (GeV) &  ${4\pi \over g_{a_1, K_a}^2}$ \\
\hline
$a_1$   &  1.190 &  2.30  & 2.25--3.30 & 0.129 & 0.404 \\
        &  1.230 &  2.45  & 2.25--3.55 & 0.132 & 0.405 \\
        &  1.270 &  2.65  & 2.25--3.80 & 0.134 & 0.417 \\
\hline
$K_a$   &  1.340 &  2.90  & 1.00--4.05 & 0.142 & 0.395  \\
\end{tabular}
\end{center}
\end{table}

\newpage
\begin{table}
\caption{Case IV : $\langle\bar{q}q\rangle$~=~--(0.230 GeV)$^3$,
$\langle\bar{s}s\rangle$ = 0.8 $\langle\bar{q}q\rangle$,
and including a mixed condensate with variation of $\alpha_s$(M).}
\label{table_d7}
\begin{center}
\begin{tabular}{ c c c c c c}
axial meson & mass (GeV)& $s_0$ (GeV$^2$) & $M^2$ (GeV$^2$) &
 $f_\pi$, $f_K$ (GeV) &  ${4\pi \over g_{a_1, K_a}^2}$ \\
\hline
$a_1$   &  1.190 &  2.30  & 1.35--3.30 & 0.128 & 0.406 \\
        &  1.230 &  2.50  & 1.35--3.60 & 0.130 & 0.420 \\
        &  1.270 &  2.70  & 1.35--3.90 & 0.132 & 0.431 \\
\hline
$K_a$   &  1.340 &  2.85  & 1.05--4.00 & 0.144 & 0.381  \\
\end{tabular}
\end{center}
\end{table}

\begin{table}
\caption{Case IV : $\langle\bar{q}q\rangle$~=~--(0.250 GeV)$^3$,
$\langle\bar{s}s\rangle$ = 0.8 $\langle\bar{q}q\rangle$,
and including a mixed condensate with variation of $\alpha_s$(M).}
\label{table_d8}
\begin{center}
\begin{tabular}{ c c c c c c}
axial meson & mass (GeV)& $s_0$ (GeV$^2$) & $M^2$ (GeV$^2$) &
 $f_\pi$, $f_K$ (GeV) &  ${4\pi \over g_{a_1, K_a}^2}$ \\
\hline
$a_1$   &  1.190 &  2.30  & 2.25--3.30 & 0.129 & 0.404 \\
        &  1.230 &  2.45  & 2.25--3.55 & 0.132 & 0.405 \\
        &  1.270 &  2.65  & 2.25--3.80 & 0.134 & 0.417 \\
\hline
$K_a$   &  1.340 &  2.85  & 1.20--4.00 & 0.144 & 0.382  \\
\end{tabular}
\end{center}
\end{table}

\newpage
\begin{table}
\caption{$f_K/f_\pi$ ratio and couplings for various cases
($\langle\bar{q}q\rangle$~=~--(0.230 GeV)$^3$).}
\label{table_d9}
\begin{center}
\begin{tabular}{ l c c c }
       &        &   \multicolumn{2}{c}{couplings} \\ \cline{3-4}
 cases &  ratio & $\frac{4\pi}{g_{a_1}^2}$ &  $\frac{4\pi}{g_{K_a}^2}$ \\
\hline
Case I : $\langle\bar{s}s\rangle=0.8\langle\bar{q}q\rangle$
&   1.08$\pm$0.02   & 0.43$\pm$0.02 &  0.43 \\
Case II : $\langle\bar{s}s\rangle=0.8\langle\bar{q}q\rangle$ + mixed con.
                  &   1.11$\pm$0.02   & 0.43$\pm$0.02 &  0.40 \\
Case III : $\langle\bar{s}s\rangle=0.8\langle\bar{q}q\rangle$ + $\alpha_s$(M)
                  &   1.09$\pm$0.02   & 0.42$\pm$0.02 &  0.40 \\
Case IV : $\langle\bar{s}s\rangle=
0.8\langle\bar{q}q\rangle$ + mixed con. + $\alpha_s$(M)
                  &   1.11$\pm$0.02   & 0.42$\pm$0.02 &  0.38 \\
\end{tabular}
\end{center}
\end{table}

\begin{table}[b]
\caption{$f_K/f_\pi$ ratio and couplings for various cases
($\langle\bar{q}q\rangle$~=~--(0.250 GeV)$^3$).}
\label{table_d10}
\begin{center}
\begin{tabular}{ l c c c }
       &        &   \multicolumn{2}{c}{couplings} \\  \cline{3-4}
 cases &  ratio & $\frac{4\pi}{g_{a_1}^2}$ &  $\frac{4\pi}{g_{K_a}^2}$ \\
\hline
Case I : $\langle\bar{s}s\rangle=0.8\langle\bar{q}q\rangle$
                  &   1.06$\pm$0.02   & 0.40$\pm$0.02 &  0.42 \\
Case II : $\langle\bar{s}s\rangle=0.8\langle\bar{q}q\rangle$ + mixed con.
                  &   1.09$\pm$0.02   & 0.40$\pm$0.02 &  0.39 \\
Case III : $\langle\bar{s}s\rangle=0.8\langle\bar{q}q\rangle$ + $\alpha_s$(M)
                  &   1.08$\pm$0.02   & 0.41$\pm$0.02 &  0.40 \\
Case IV : $\langle\bar{s}s\rangle=
0.8\langle\bar{q}q\rangle$ + mixed con. + $\alpha_s$(M)
                  &   1.09$\pm$0.02   & 0.41$\pm$0.02 &  0.38 \\
\end{tabular}
\end{center}
\end{table}

\newpage
\begin{table}
\caption{$f_K/f_\pi$ ratio from various models}
\label{table_d11}
\begin{center}
\begin{tabular}{l l c}
      models    &     ratio   &    references \\
\hline
 Experiment		      & 1.22~$\pm$~0.02  &  Ref. \cite{PRD} \\
 SU(3) $\sigma$ Model	      & 1.31		 &  Ref. \cite{Ch}  \\
 Current Algebra	      & 1.3~$\pm$~0.3	 &  Ref. \cite{Do} \\
 Chiral Perturbation Theory   & 1.21		 &  Ref. \cite{La} \\
 Bag Model I		      & 1.02		 &  Ref. \cite{Go}\\
 Bag Model II		      & 1.2		 &  Ref. \cite{Ha}\\
 Potential Model I	      & 1.27		 &  Ref. \cite{Kr}\\
 Potential Model II	      & 1.27		 &  Ref. \cite{God}\\
 Relativistic Quark Model     & 1.25		 &  Ref. \cite{Ko} \\
 Finite Q$^2$ Sum Rule (FQSR) & 1.15		 &  Ref. \cite{Sa} \\
 Effective Lagrangian Method I	& 1.15		 &  Ref. \cite{Mc} \\
 Effective Lagrangian Method II & 1.13		 &  Ref. \cite{Mu} \\
 Electroweak Theory I	      & 1.22~$\pm$~0.01  &  Ref. \cite{Le} \\
 Electroweak Theory II	      & 1.23~$\pm$~0.02  &  Ref. \cite{Ho} \\
 Lattice Gauge Theory I       & 1.35		 &  Ref. \cite{De}\\
 Lattice Gauge Theory II      & 1.10		 &  Ref. \cite{Ham}\\
 Lattice Gauge Theory III     & 1.16~$\pm$~0.07  &  Ref. \cite{Ab}\\
 Lattice Gauge Theory IV      & 1.16		 &  Ref. \cite{Bu}\\
 Nambu--Jona-Lasinio Model I  & 1.03		 &  Ref. \cite{Kl}\\
 Nambu--Jona-Lasinio Model II & 1.05		 &  Ref. \cite{Hat}\\
 Nambu--Jona-Lasinio Model III & 1.12             &  Ref. \cite{Al}\\
 QCD Sum Rules I	      & 1.22		 &  Ref. \cite{Re}\\
 QCD Sum Rules II$^a$	      & 1.11~$\pm$~0.02  &  present work  \\
\end{tabular}
\end{center}
$^a$ Case IV and $\langle\bar{q}q\rangle$ = --(0.230 GeV)$^3$ (see text).
\end{table}

\end{document}


\draft

\centerline{\Large Erratum : $f_K/f_\pi$ ratio from QCD sum rules} 

\vspace{1cm}

In this erratum we correct an error in the previous calculation
and present a more detailed analysis including the variations of
the decay constants $f_\pi$ and $f_K$ on input parameters.

In the case of the $K_a$ sum rule, there is no contribution
of the mixed condensate, i.e.,
$m_s \langle g_s \bar{s} ~\sigma \cdot G ~s\rangle$.  Therefore, the argument
on the mixed condensate in Sec. III B and D becomes irrelevant.  
When we include the effects from the anomalous dimension, 
Eq. (10) 
is rewritten as
\begin{eqnarray}
f_{\pi,K}^2 e^{-\frac{m_{\pi,K}^2}{M^2}}
&+& {m_{a_1,K_a}^2 \over g_{a_1,K_a}^2} e^{-\frac{m_{a_1,K_a}^2}{M^2}}
\nonumber\\*
&=& {1 \over 4\pi^2} M^2
\left[ \left( 1+\frac{\alpha_s(M)}{\pi} \right)
\left( 1-e^{-\frac{s_0}{M^2}} \right)
+\frac{A}{M^2} L^{-{8\over 9}}
+\frac{B}{M^4}
+\frac{C}{M^6} L^{-{1\over 9}} \right] ,
\label{form2}
\end{eqnarray}
where $L$ is defined by
\begin{eqnarray}
L \equiv \frac{\alpha_s (\mu)}{\alpha_s (M)}
\end{eqnarray}
and
\begin{eqnarray}
\alpha_s (M) = \frac{4\pi}{9} \frac{1}{ln(M^2/\Lambda^2)} .
\end{eqnarray}
Here, $\mu$ is the renormalization point taken to be 1 GeV
and $\Lambda$ the QCD scale parameter, 0.25 GeV.

Using the  methods described in the paper,
we get 0.133 GeV and 0.145 GeV for  $f_\pi$ and $f_K$, respectively,
for the basic input values given 
in Tables \ref{table_fpi} and \ref{table_fk}.
We also include variations of these decay constants for
other inputs, 
which are coming from the uncertainty of the basic inputs as discussed in 
the paper.
 For example, the first line in Table \ref{table_fpi} shows that
$f_\pi$ = 0.131 GeV (or 0.135 GeV) if we change the quark condensate to 
$\langle\bar{q}q\rangle$ =  --(0.210 GeV)$^3$ (or --(0.250 GeV)$^3$)
while other basic inputs are fixed. 
$f_K$ is most sensitive to the strange quark mass $m_s$,
while $f_\pi$ to the $a_1$ mass.
Including the variations, we get the maximum ratio of
$f_K$/$f_\pi$ = 1.16.
Here, we do not include the variation of $f_K$
on  $m_{K_a}$.

Some references use the renormalization point
 $\mu$ = 0.5 GeV and the QCD scale parameter $\Lambda$ = 0.15 GeV.
If we take these values of  $\mu$ and $\Lambda$, we get 0.131 GeV and
 0.145 GeV for   $f_\pi$ and $f_K$, respectively, for the same basic inputs.
In the tables we also present the variations coming from these changes.

\begin{table}
\caption{$f_\pi$ and its variations. 
Other inputs means  other possible inputs discussed 
in the paper. 
[$\cdots$] corresponds
to taking $\mu$ = 0.5 GeV. }
\label{table_fpi}
\begin{center}
\begin{tabular}{ c c c}
 basic inputs &  other inputs  &  variations (MeV) \\
\hline
$\langle\bar{q}q\rangle$ =  --(0.230 GeV)$^3$             &
--(0.210 GeV)$^3$, --(0.250 GeV)$^3$       &  $\mp$ 2 [$\mp$ 3]     \\
$\langle\frac{\alpha_s}{\pi} G^2\rangle $ = 0.015 GeV$^4$  &
                            0.012 GeV$^4$  &   --~ 2  [--~ 2]          \\
$\Lambda$     =    0.25 [0.15] GeV                              &
                            0.20 [0.10] GeV  &   --~ 2  [--~ 3]    \\
$m_{a_1}$         =    1.230 GeV                              &
                 1.190, 1.270 GeV      &  $\mp$ 3   [$\mp$ 3]          \\
\end{tabular}
\end{center}
\end{table}

\begin{table}
\caption{$f_K$ and its variations.  
Other inputs means  other possible inputs discussed 
in the paper. 
[$\cdots$] corresponds
to taking $\mu$ = 0.5 GeV.}
\label{table_fk}
\begin{center}
\begin{tabular}{c c c}
 basic inputs &  other inputs  &  variations (MeV) \\
\hline
$\langle\bar{q}q\rangle$ =  --(0.230 GeV)$^3$             &
--(0.210 GeV)$^3$, --(0.250 GeV)$^3$       &  $\pm$ 1   [$\pm$ 1]         \\
$\langle\frac{\alpha_s}{\pi} G^2\rangle $ = 0.015 GeV$^4$  &
                            0.012 GeV$^4$  &   --~ 2  [--~ 2]           \\
$\Lambda$     =    0.25 [0.15] GeV                               &
                         0.20 [0.10] GeV   &   --~ 3  [--~ 4]            \\
$m_{K_a}$         =    1.340 GeV                              &
                 1.270, 1.400 GeV      &  $\mp$ 4  [$\mp$ 4]           \\
$\langle\bar{s}s\rangle$ = 0.8 $\langle\bar{q}q\rangle$   &
      0.7, 0.9  $\langle\bar{q}q\rangle$   &  $\pm$ 1  [$\pm$ 1]          \\
$m_s$ = 0.150 GeV                                         &
                0.120, 0.180 GeV       &  $\pm$ 7    [$\pm$ 6]        \\
\end{tabular}
\end{center}
\end{table}
